# Observation and analysis of the new W-type W UMa eclipsing binary VSX J053024.8+842243

D. R. S. Boyd

*West Challow, Wantage, Oxfordshire, OX12 9TX, UK [davidboyd@orion.me.uk]*

**ABSTRACT**

Using multicolour photometry we have confirmed the binary nature of the new W-type W UMa eclipsing binary VSX J053024.8+842243 and established its primary eclipse ephemeris to be HJD = 2455924.38150(26) + 0.4322929(1) * E. Using the light curve modelling code PHOEBE and published data on the evolution of W-type contact binaries we found the primary and secondary components to have masses 0.50 $M_\odot$ and 1.44 $M_\odot$, radii 0.87 $R_\odot$ and 1.42 $R_\odot$, luminosities 0.98 $L_\odot$ and 1.91 $L_\odot$, temperatures 6145 K and 5702 K and binary orbit inclination 59.4°. We found the distance to the binary to be 511 parsec, its E(B-V) colour excess 0.04 and its intrinsic (B-V) colour index 0.62. A low resolution spectrum corrected for interstellar reddening confirmed its spectral type as G2V.

## 1 INTRODUCTION

Sebastian Otero compiled a list of suspected eclipsing binaries on the basis of survey observations (Otero 2011). One of the stars in his list, GSC 04621-00193 (= 2MASS J05302483+8422431), is located in the constellation of Cepheus at RA 05h 30m 24.8s Dec +84° 22' 43" (J2000). Its suspected orbital period was listed as 0.355 days. A search of survey databases revealed data on the star in The Amateur Sky Survey (Droege et al. 2006) and the Northern Sky Variability Survey (Wozniak et al. 2004). The TASS data was noisy and revealed no clearly defined periods. The NSVS data, recorded over a 20 week period in 1999, was of higher quality and a period search using Peranso (Peranso 2015) gave a clearly defined period at 0.355 days apparently confirming the suggested period. However these were small datasets of 92 and 82 observations respectively. Accurate and more extensive photometry would be necessary to verify its true nature.

## 2 PHOTOMETRIC OBSERVATION

V-band CCD photometry was carried out on 3 nights in December 2011 using a 0.35m SCT and SXVR-H9 CCD camera. This produced 1200 images. As this was a new variable not in any of the comparison star databases, suitable comparison stars had to be identified. Johnson B and V and Sloan g, r and i magnitudes for nearby stars were found from the AAVSO Photometric All-Sky Survey (APASS 2015) and conversion formulae from the Sloan Digital Sky Survey website (SDSS 2015) were used to calculate Cousins Rc and Ic magnitudes. Five stars close to the variable were chosen as comparison stars. All images were dark-subtracted and flat-fielded and ensemble differential aperture photometry performed with respect to these five comparison stars to find V magnitudes for the variable. Further observations using B, V, Rc and Ic filters were made on 5 nights in July and August 2012 and the derived magnitudes were transformed onto the standard photometric system using the procedure described in Boyd (2012). Average magnitude uncertainties were 0.018 in B, 0.008 in V, 0.011 in Rc and 0.021 in Ic. Observations were obtained on two further nights in December 2015 and January 2016 to refine the orbital period. The times of all observations were converted to Heliocentric Julian Dates (HJD/UTC) to correct for the orbital motion of the Earth.

## 3 ORBITAL EPHEMERIS DETERMINATION

Period analysis of the 2011 V-band data plus the NSVS data from 1999 showed a signal at 0.3553(5) days but a stronger signal at 0.4323(5) days. These signals differ in frequency by 0.5 cycles/day, not unexpected for a W UMa binary. Inspection of the light curve phased on these periods immediately revealed that the longer one was the correct orbital period. Figure 1 shows the Peranso plot of these data phased on the 0.4323 d orbital period.

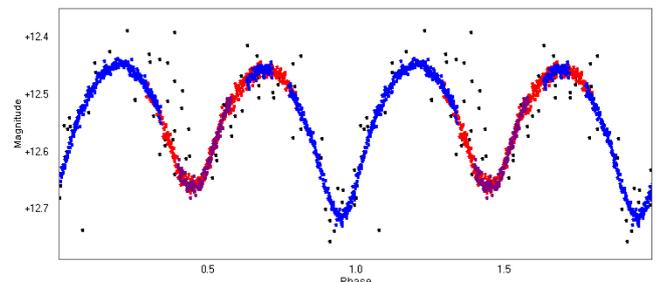

**Figure 1**: Plot of V-band data from 2011 and NSVS data from 1999 phased on a 0.4323 day period showing two complete cycles with colours indicating 2011 data recorded on different dates. The sparse NSVS data recorded over a 20 week period in 1999 are shown in black.

Times of minimum of primary eclipses in 2011 and 2012 were obtained by a quadratic fit to the lower part of the eclipse light curve. Using the above period a cycle number was assigned to each eclipse and an improved orbital period calculated. Based on this evidence and the V-band phase diagram, the star was accepted as a new W UMa eclipsing binary for inclusion in the AAVSO Variable Star Index (VSX 2015) with the designation VSX J053024.8+842243 (henceforth VSX053024) and variability type EW following the definition in the General Catalogue of Variable Stars (GCVS 2015).

Including further primary eclipses observed in 2015 and 2016, the following linear ephemeris for the times of primary eclipse minima was determined with E as the cycle number and the figures in brackets representing the uncertainties in the last digits:

$T_0$ (HJD/UTC) = 2455924.38150(26) + 0.4322929(1) * E

## 4 BINARY PARAMETER ANALYSIS

Using this ephemeris and the multicolour observations from 2012, phased orbital light curves in B, V, Rc and Ic were generated (Figure 2). Analysis of the light curves to determine physical parameters of the binary system was carried out using the program PHOEBE (PHysics Of Eclipsing BinariEs), ver. 0.31a (Prsa & Zwitter 2005, PHOEBE 2015). PHOEBE is based on the Wilson-Devinney algorithm (Wilson & Devinney 1971) and its later modifications. A model of the binary system is computed iteratively from an initial set of user-defined parameter values and synthesised light curves generated from the model are fitted to the observed phased light curves. A cost function computed as the weighted $\chi^2$ discrepancy between the synthesised light curves and the observed data is used to evaluate the quality of the solution.

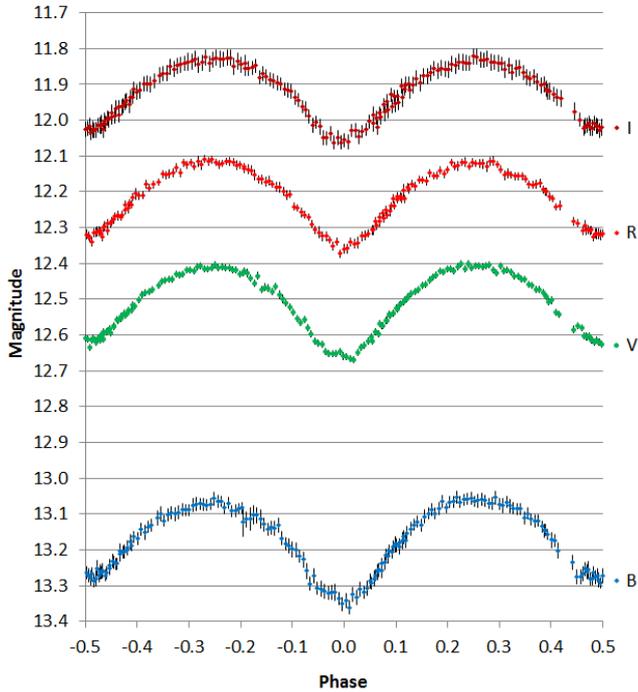

**Figure 2**: Light curves in B, V, Rc and Ic wavebands phased on the orbital period of 0.4322929 days.

The depths of the primary and secondary minima in the orbital light curve are different indicating that the temperatures of the two stars, although similar, are not the same. For this reason, attempting to model the system as a W UMa type with the components in thermal contact failed. Therefore the PHOEBE mode *Overcontact binary not in thermal contact* was used. The primary or deepest minimum is conventionally taken as phase 0.0. Since the secondary eclipse occurs exactly at phase 0.5, and as we shall see the components are transferring mass, the orbit was assumed to be circular.

The mean apparent colour index of the star at quadrature (phases 0.25 and 0.75) when both components are fully visible is (B-V) = 0.66±0.01. Interpolating Table 5 in Pecaut & Mamajek (2013), this is equivalent to an effective temperature of 5721 K. Although the statistical uncertainty in this mapping is around ±20 K the real uncertainty is larger, possibly ±100 K, because of the effect of other factors such as metallicity. Initially we assigned 5721 K as the effective temperature of the primary T1 but recognised that this might have to be revised once the temperature difference between the components had been determined. No correction for interstellar reddening was made at this stage as the distance to the star was not known.

The epoch of the primary minimum and the orbital period were taken from the ephemeris. Conventionally the primary is the star which is eclipsed at the primary minimum and is therefore the brighter star. Since stars with this temperature have convective atmospheres, the bolometric albedo of both stars was set to 0.5 (Rucinski 1969) and the gravity darkening coefficients to 0.32 (Lucy 1967). Limb darkening was determined using a logarithmic law with bolometric coefficients interpolated from van Hamme's tables (van Hamme 1993).

In the absence of radial velocity measurements, the conventional way of determining the stellar mass ratio q = M2/M1 is the q-search method. A specific value was assigned to q and the other parameters were allowed to vary, specifically the effective temperature of the secondary T2, the orbital inclination i, the modified Kopal potential (Kopal 1959) of the primary Ω1 and the relative luminosities of the primary in each colour band. The PHOEBE solution was iterated until it stabilised and the value of the cost function was noted. The value of q was varied over a wide range, the solution repeated each time and the resulting cost functions plotted against q. In the q-search method, the minimum of the cost function is assumed to indicate the most likely value for q. Figure 3 shows how the cost function varied with q and includes an enlargement of the region around the lowest minimum. Larger values of q than 3.1 produced solutions which PHOEBE reported as unphysical. We then took q = 2.85 at the cost function minimum as the starting value and repeated the solution, this time with q variable, until it stabilised. The resulting value was q = 2.855 and this was then fixed.

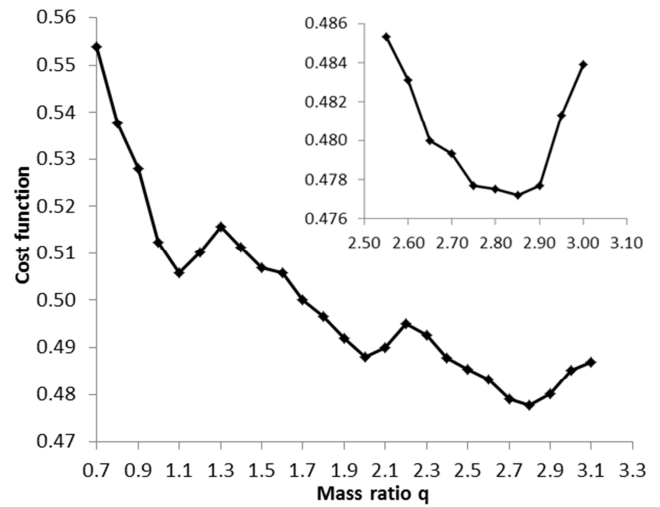

**Figure 3**: Variation of the PHOEBE cost function with mass ratio q showing the region around the lowest minimum at a larger scale.

The PHOBE solution indicated that the temperature of the secondary was cooler than the primary, as expected, while the mass and radius of the secondary were larger than those of the primary. According to the definition in Binnendijk (1970), this is a W-type of W UMa contact binary.

The relatively large temperature difference of 370 K between the components indicated that we had underestimated the temperature of the primary based on the observed value of (B-V). Using the temperatures, radii and luminosities of the components from the PHOBE solution, we calculated the effective temperature which we would expect to observe for the binary seen as a single object and found this to be 5460 K, clearly much lower than the temperature indicated by our (B-V) measurement. We therefore adjusted our estimate of the primary temperature and reran the analysis until the temperature we would expect to observe matched our (B-V) measurement. This gave revised primary and secondary effective temperatures of 6010 K and 5598 K, again with no allowance for any potential reddening.

Without radial velocity data we have no absolute measurement of the spatial scale of the system, in particular of its semi-major axis, so we adopted a heuristic approach to determining this. Harmanec (1988) provides a formula, based on analysis of published data on detached main sequence stars, for calculating the mass of a main sequence star from its effective temperature. Using this formula and the effective temperatures of the two stars derived above we calculated values for their masses. As this formula is for detached main sequence stars it will not accurately predict the current masses of the components of close contact binaries which are filling or almost filling their Roche lobes and therefore likely to be exchanging mass. However it is reasonable to assume that it will provide a useable estimate of the initial masses of component stars with those effective temperatures prior to the onset of mass transfer.

Yildiz & Dogan (2013) tabulate the parameters of 49 W-type W UMa binaries and give estimated initial masses of these stars obtained by modelling changes in the masses of the components due to mass transfer between the components and mass loss from the binary system. Note that they adopted the convention that the currently more massive star is the primary so we have exchanged their primary and secondary stars for use here. They propose that the star which we are calling the primary as it is hotter and more luminous was originally the more massive star (q < 1) but that it subsequently transferred matter to the secondary to the point where the situation has reversed and now q > 1. They argue that the current luminosity of the stars is determined principally by their initial rather than their current masses.

We plot in Figure 4 the initial vs current masses of the primary and secondary stars of the W-type contact binaries tabulated in Yildiz & Dogan (2013) along with linear fits to these distributions which represent average evolutionary trajectories from initial to current masses. This trajectory is more tightly defined for the secondary than for the primary. We used these evolutionary trajectories to convert the initial masses calculated from the Harmanec (1988) formula to possible current masses for the two stars. These are also marked in Figure 4 and are within the spread of other W-type contact binaries.

Using these current masses, M1 and M2, and the orbital period, P, we calculated a preliminary estimate for the semi-major axis of the binary as a = 3.01 $R_\odot$ using Kepler's third law:

$$4\pi^2 a^3 = G(M1+M2)P^2$$

We then stepped the semi-major axis through a series of values around this region and used PHOEBE to calculate corresponding component masses each time. Our aim was to find the value of the semi-major axis which gave closest agreement between component masses calculated by PHOEBE and those derived via the above evolutionary mapping. We did this by finding the value of the semi-major axis which minimised the distance in the M1 vs M2 plane between the masses calculated by PHOEBE and the masses obtained from the evolutionary mapping. Figure 5 shows the current masses derived from the mapping and the trajectory in the M1 vs M2 plane of the component masses calculated by PHOEBE as the semi-major axis was varied. The closest agreement occurs with a = 2.98 and this was adopted as the semi-major axis of the binary.

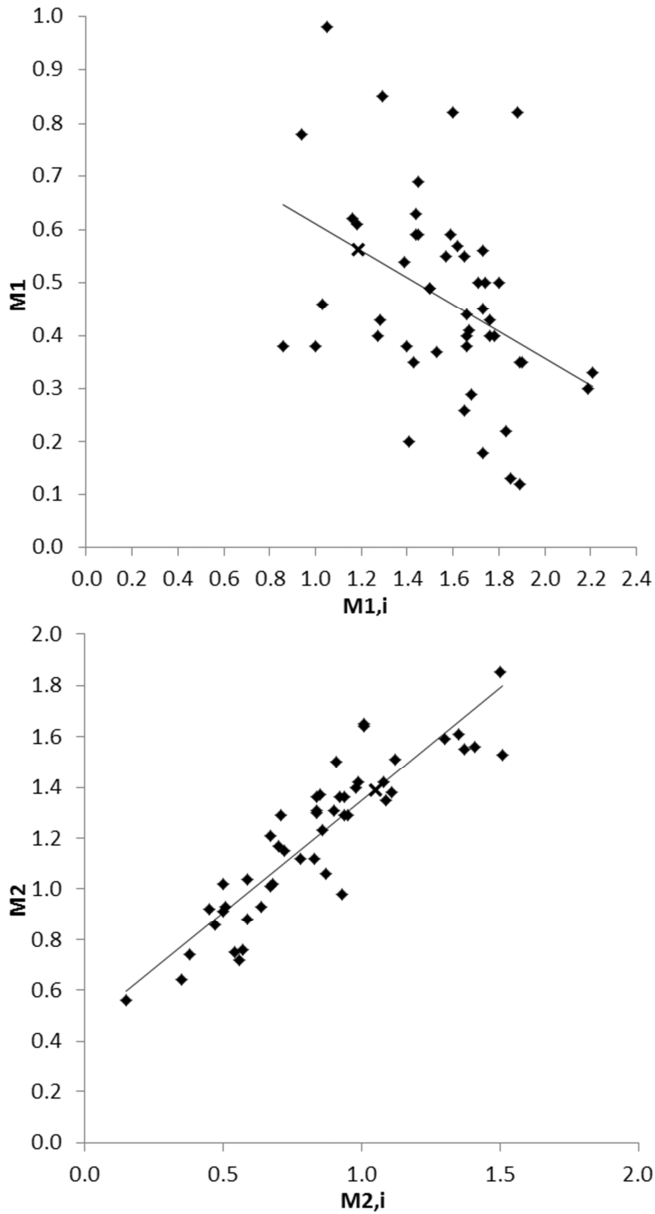

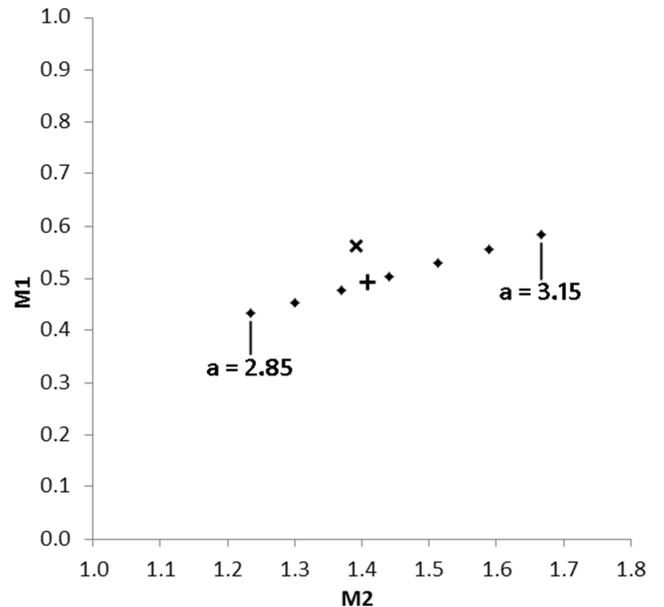

**Figure 5**: The location of the current component masses derived from the evolutionary trajectories in Figure 4 is marked x and the path of the component masses calculated by PHOEBE as the semi-major axis, a, was varied between the values indicated is marked by diamonds. The location of the adopted component masses at a = 2.98 is marked +.

Fixing the component effective temperatures and the mass ratio, inclination and semi-major axis we used PHOEBE to calculate the masses, radii, bolometric magnitudes and surface gravities of the two components of VSX053024. We used the Stefan-Boltzmann Law to compute the luminosities of the stars from their radii and temperatures. These parameter values, which we refer to as interim at this stage, are listed in Table 1. The uncertainties on T2, q and i are formal errors from the fit and do not include systematic errors arising from assumptions made in the model fitting procedure so are likely to be underestimates. In view of the heuristic way in which the semi-major axis and hence the masses, radii, luminosities, bolometric magnitudes and surface gravities were calculated, we do not give uncertainties for these.

**Figure 4**: Plots of initial (Mx,i) vs current (Mx) masses of the primary (top) and secondary (bottom) of 49 W-type contact binaries tabulated in Yildiz & Dogan (2013) (with components reversed as noted in the text) with linear fits representing average evolutionary trajectories from initial to current masses. The primary and secondary components of VSX053024 with initial masses from the formula in Harmanec (1988) and current masses lying on the average mappings are marked x.

To investigate our confidence in the value of q determined from q-search, we plot in Figure 6 the masses and radii of the two components of the W-type contact binaries tabulated in Yildiz & Dogan (2013) plus the masses and radii which the two components of VSX053024 would have using the values of q at each of the three minima in Figure 3. It is clear that the final value of q at the deepest minimum is most consistent with other W-type binaries.

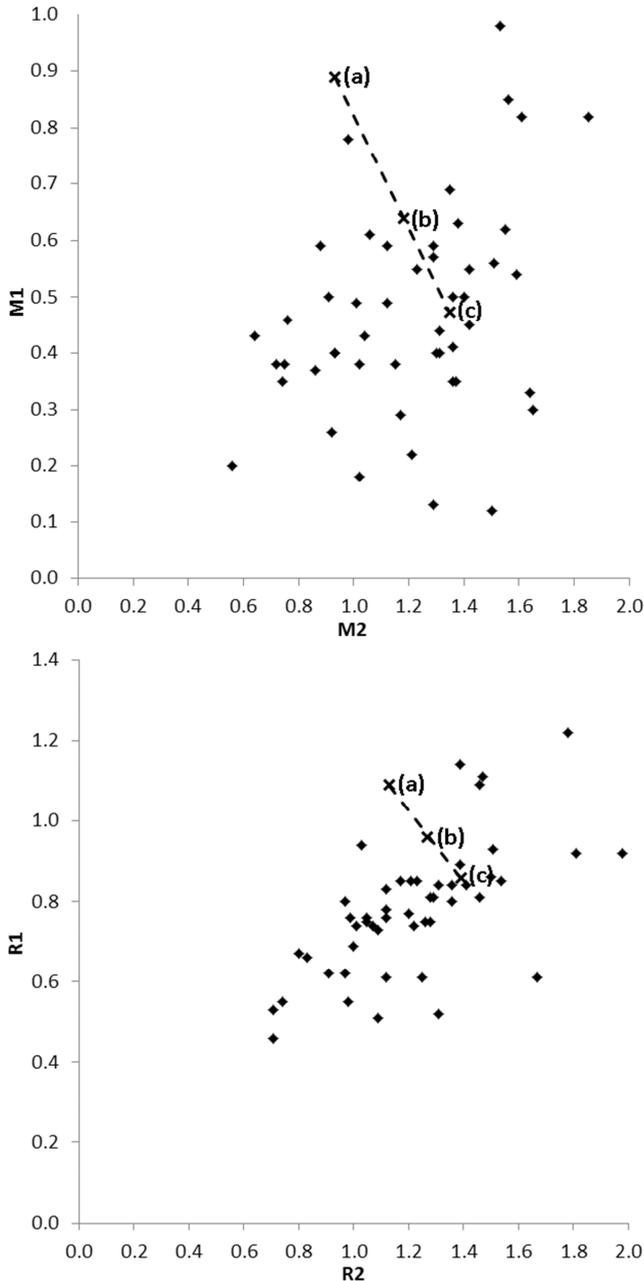

**Figure 6**: Distributions of the masses (top) and radii (bottom) of the two components of 49 W-type contact binaries tabulated in Yildiz & Dogan (2013) (with components reversed as noted in the text) with dotted lines showing the trajectories of the masses and radii of the components of VSX053024 during q-search with (a) q = 1.15, (b) q = 1.92, (c) q = 2.855 corresponding to the three minima in Figure 3.

Figure 7 shows mass-radius and mass-luminosity diagrams for the W-type contact binaries from Yildiz & Dogan (2013) with lines representing the Zero Age Main Sequence (ZAMS) from Tout et al. (1996) assuming metallicity Z = 0.02. The locations of the two components of VSX053024 are marked. The primary is now considerably less massive that its equivalent radius on the main sequence would imply but it has maintained approximately its original luminosity, consistent with the model of mass loss proposed by Yildiz & Dogan (2013). The secondary remains close to the main sequence. The location of the components of VSX053024 in these diagrams is broadly consistent with the distribution of primary and secondary stars for other W-type contact binaries.

**Table 1**: Binary system parameter values for VSX053024. Interim values are before the correction for reddening, final values are after the correction for reddening.

| Parameter | Interim value | Final value |
| --- | --- | --- |
| P [d] | 0.4322929 | 0.4322929 |
| $T_0$ [HJD] | 2455924.38150 | 2455924.38150 |
| T1 [K] | 6010 (fixed) | 6145 (fixed) |
| T2 [K] | 5598±10 | 5702±10 |
| q = M2/M1 | 2.855±0.003 | 2.855±0.003 |
| i [°] | 59.49±0.09 | 59.43±0.09 |
| a [$R_\odot$] | 2.98 | 3.00 |
| M1 [$M_\odot$] | 0.49 | 0.50 |
| M2 [$M_\odot$] | 1.41 | 1.44 |
| R1 [$R_\odot$] | 0.87 | 0.87 |
| R2 [$R_\odot$] | 1.41 | 1.42 |
| L1 [$L_\odot$] | 0.89 | 0.98 |
| L2 [$L_\odot$] | 1.75 | 1.91 |
| Mbol1 | 4.18 | 4.08 |
| Mbol2 | 4.92 | 4.81 |
| log g1 | 4.25 | 4.26 |
| log g2 | 4.29 | 4.29 |
| Mv1 | 4.27 | 4.18 |
| Mv2 | 5.06 | 4.93 |
| Mv | 3.84 | 3.74 |
| Av | 0.129 | 0.135 |
| E(B-V) | 0.042 | 0.044 |
| D [pc] | 488 | 511 |
| $(B-V)_0$ | 0.62 | 0.62 |

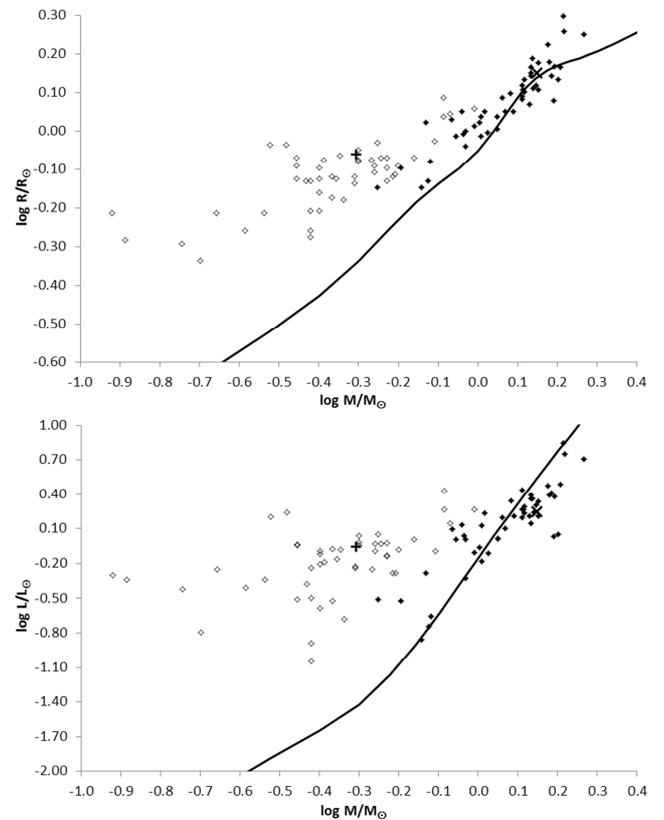

**Figure 7**: Mass-radius (top) and mass-luminosity (bottom) diagrams of the components of 49 W-type contact binaries from Yildiz & Dogan (2013) (with components reversed as noted in the text). Open and solid symbols represent the primary (less massive) and secondary (more massive) components respectively. The primary and secondary components of VSX053024 as calculated by PHOEBE are shown as + and x respectively. The solid line represents the ZAMS trajectory from Tout et al. (1996).

# 5 ESTIMATING THE DISTANCE TO THE STAR

According to the NASA IPAC database (NASA 2015), interstellar extinction and reddening due to galactic dust between the Earth and the edge of the galaxy along the line of sight towards the star are Av = 0.208 and E(B-V) = 0.067. However, only when we know the distance to the star can we say what values for extinction and reddening it is reasonable to adopt for the star itself.

Using the bolometric magnitude of each star, Mbol1 and Mbol2, obtained from PHOEBE and the V band bolometric corrections from Pecaut & Mamajek (2013) appropriate to the effective temperatures of the primary and secondary, the absolute V magnitude Mv of each star was calculated from

$$Mv = Mbol - BCv$$

The absolute V magnitude of the binary system was then found by adding the equivalent fluxes of the two components. Using this absolute V magnitude, the apparent V magnitude of the star at quadrature V = 12.41 found from its light curve, and taking the V band extinction between Earth and the star as Av = 0.208, an initial estimate of the distance d to the star was obtained from the relation

$$5 \log_{10}(d) = 5 + V - Mv - Av$$

This gave the distance as 470 parsec. Given the star's galactic coordinates of l = 129° and b = 25°, it is approximately 62% of the way to the edge of the galactic thin disk (Juric et al. 2008). We therefore scaled the values from the NASA IPAC database accordingly to get interstellar extinction and reddening to the star of Av = 0.129 and E(B-V) = 0.042. Using this value of Av gave an improved distance estimate of 488 parsec. Given the uncertainty in this derivation of the extinction to the star we did not consider further iteration of this distance calculation worthwhile. Using the relation

$$(B-V)_0 = (B-V) - E(B-V)$$

with the apparent colour index (B-V) = 0.66±0.01 from the light curves and the colour excess E(B-V) = 0.04 gave the intrinsic colour index of the star $(B-V)_0$ = 0.62±0.01.

# 6 FINAL ITERATION

This value of the intrinsic colour index corrected for interstellar reddening indicated that the current temperatures for the components were underestimates. Again using Table 5 in Pecaut & Mamajek (2013) with an intrinsic colour index of 0.62, this indicated an effective temperature of 5834K for both components seen as a single object. The above analysis was repeated to find a solution which was consistent with this temperature. Final values for all parameters are given in Table 1. As the effect of these changes on Figures 4, 5, 6 and 7 was small and did not change the information they convey, we do not repeat these plots here with the new parameter values. The change in interstellar reddening from the revised distance of 511 parsec was sufficiently small that it did not alter the intrinsic colour index of the star given above so at this point we considered the analysis finished.

A comparison between the measured and synthesised orbital light curves is shown in Figure 8. The heights of the two maxima in the light curves are very similar indicating that there are unlikely to be any major spotted regions on either component. We experimented with adding starspots but found that while marginal improvements could be made to the fit in individual colours, this usually resulted in the fits for other colours becoming worse. In the end we decided to leave the stars spotless following the advice in the PHOEBE Science Manual that "One should thus *never* (sic) add spots to the model if there is no firm evidence that points to it, e.g. accurate Doppler tomography based on high-resolution spectroscopy". Figure 9 presents schematic representations of the relative shapes and sizes of the two stars at phases 0.00 and 0.25 showing their significant ellipsoidal distortion.

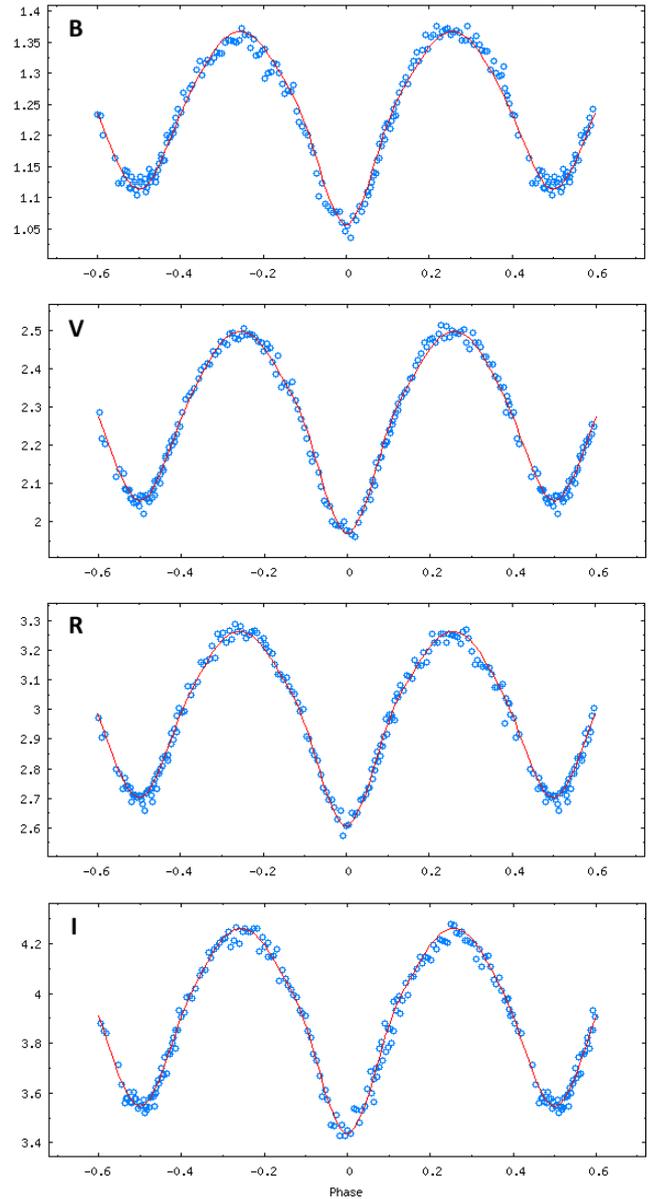

**Figure 8**: Comparison of measured (blue circles) and synthesised (red lines) orbital light curves for B, V, Rc and Ic wavebands.

# 7 SPECTROSCOPIC OBSERVATION

As a final check on the consistency of our analysis, fourteen low resolution (R=1200) spectral images of the binary were obtained in December 2015 using a LISA spectrograph on a 0.28m SCT. At the time the binary orbit was at phase 0.26 so both components were fully visible. The spectral images were dark-subtracted, flat-fielded and wavelength calibrated using Ar-Ne calibration lamp spectra obtained before and after the stellar spectra. Corrections for atmospheric and instrumental response were made using spectra of the A4V star HD4853 observed at the same altitude as the binary. The spectra of the binary were combined and flux calibrated using a concurrently measured V magnitude for the star according to the procedure described in Boyd (2014).

The binary spectrum was then corrected for extinction and reddening using Av = 0.135, E(B-V) = 0.044 calculated using the revised distance and the formulae for normalised extinction Aλ/Av given by Cardelli et al. (1989). By comparing this final spectrum with spectra from the Pickles Stellar Spectral Flux Library (Pickles 1998, Pickles 2015), we found the closest visual match to a standard spectral subtype of the MK Spectral Classification System (Keenan 1984) to be G2V. Although not a formal spectral classification using line strengths, this subjective assessment was sufficient for our purpose. It was consistent with our finding that the component stars remain close to the main sequence. Figure 10 shows the spectrum smoothed with a Gaussian of order 3 together with the Pickles G2V spectrum. According to Gray & Corbally (2009) the G2V spectral type corresponds to an intrinsic colour index $(B-V)_0$ of 0.63 which is consistent with the value 0.62±0.01 derived above from the light curves.

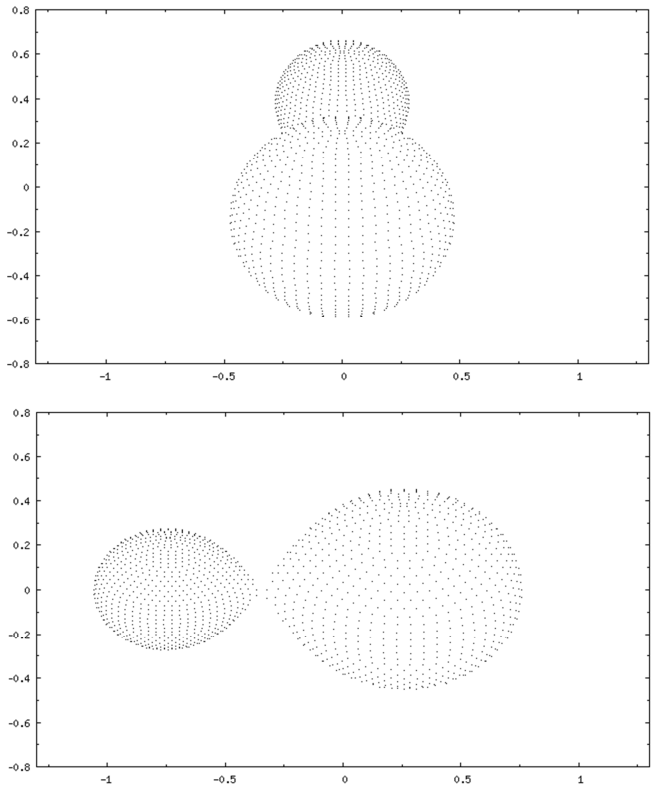

**Figure 9**: Schematic representation of the component stars of VSX053024 at phases 0.00 (top) and 0.25 (bottom). The primary hotter, less massive star is behind and on the left respectively.

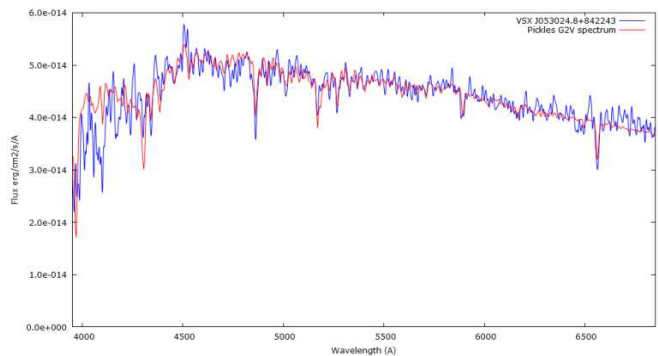

**Figure 10**: Comparison of the extinction corrected and dereddened spectrum of VSX053024 smoothed with a Gaussian of order 3 with the spectrum of a spectral type G2V star from the Pickles library.

Based on the parameters from the binary solution, the radial velocity semi-amplitudes of the two stars should be 224 and 78 km/s. While radial velocities of this size are measurable with the spectral resolution of this equipment using cross-correlation techniques, this star is too faint and the signal to noise ratio of the spectra too low for this to be practical.

# 8 CONCLUSION

We have analysed the times of eclipse minima and multicolour light curves of the new W-type W UMa eclipsing binary VSX J053024.8+842243 to determine its orbital ephemeris and the physical parameters of its component stars. In the absence of radial velocity data we have determined the absolute scale of the binary by assuming its evolutionary development has been similar to that of other W-type contact binaries and used this to determine masses, radii, luminosities and bolometric magnitudes of the component stars. From the bolometric magnitudes we estimated the distance of the binary to be 511 parsec and as a consequence its E(B-V) colour excess to be 0.04 and its intrinsic (B-V) colour index to be 0.62. A low resolution spectrum corrected for interstellar reddening indicated a spectral type of G2V and hence an intrinsic (B-V) colour index of 0.63, consistent with the photometric observations.

## ACKNOWLEDGEMENTS

I am grateful to Dr John Southworth for his advice on binary system analysis and to the referees whose constructive comments have helped to improve the paper. This research has made use of the SIMBAD database, operated at CDS, Strasbourg, France. We also acknowledge use of the AAVSO Photometric All-Sky Survey (APASS) and of the International Variable Star Index (VSX) database, operated at AAVSO, Cambridge, Massachusetts, USA.